\documentclass[11pt]{article} 
\usepackage{amsfonts, amssymb, amsmath,color, lineno,enumitem}
\usepackage[colorinlistoftodos]{todonotes}

\def\be{\begin{eqnarray}}
\def\ee{\end{eqnarray}}
\def\bee{\begin{eqnarray*}}
\def\eee{\end{eqnarray*}}
 \def\pmx{\begin{pmatrix}}
 \def\emx{\end{pmatrix}}
 \def\bsq{\begin{subequations}}
\def\esq{\end{subequations}}

 \def\zero { \begin{smallmatrix}  0 & 0 \\ 0 & 0 \end{smallmatrix} }
 \def\zero col{  \begin{smallmatrix} 0 \\ 0 \end{smallmatrix} }
  \def\zero row{  \begin{smallmatrix} 0 & 0 \end{smallmatrix} }

\newtheorem{thm}{Theorem}

\newtheorem{lemma}[thm]{Lemma}

\newtheorem{remark}[thm]{Remark}

        \def\tr{\hbox{\rm Tr} \, }
          
          \def\trp{\hbox{\rm Tr} }
     
     \def\half{{\textstyle \frac{1}{2}}}
     \def\nn{\nonumber}
\def\cA{{\cal A}}

\def\id{{\cal I}}
\def\cW{{\cal K}}
\def\cK{{\cal K}}

\def\cB{{\cal B}}

\def\cN{{\cal N}}

\def\hv{{\rm Holv}}
\def\av{{\rm Av}}

\def\eps{\epsilon}

\def\bra{\langle}
\def\ket{\rangle}
\def\kb{ \ket \bra }

\def\rt2{ \frac{1}{\sqrt{2}} }
\def\lraw{\leftrightarrow}
\def\raw{\rightarrow}
\def\lraw{\longrightarrow}

           \def\wtd{\widetilde}
\newcommand{\proj}[1]{ | #1 \kb  #1|}
\newcommand{\ovb}[1]{\overline{ #1 }}
  \def\qed{ \hfill {\rm QED}}

\def\spn{{\rm span}}

\def\ot{\otimes}
\def\op{\oplus}

   \def\av{{\rm Av}}

\title{Local Additivity Revisited}

  \author{Mary Beth Ruskai \\
  Department of Mathematics \\ University of Vermont  05405 USA \\ {\small mbruskai@gmail.com} \\ 
  \and Jon Yard \\  Institute for Quantum Computing \\ $\qquad$  Department of Combinatorics and Optimization  $\qquad$ \\
 University of Waterloo \\    Perimeter Institute for Theoretical Physics\\ Waterloo, Ontario, Canada  N2L 3G1  \\ {\small jyard@uwaterloo.ca}}

\bigskip

\begin{document}

\maketitle

\begin{abstract}
We make a number of simplifications in Gour and Friedland's proof of local additivity of 
minimum output entropy of a quantum channel.
We follow them in reframing the question as one about entanglement entropy of bipartite
states associated with a $d_B \times d_E $ matrix.  We use a different approach to reduce
the general case to that of a square positive definite matrix.
We use the integral representation of the log to  obtain  expressions for the
first and second derivatives of the entropy, and then
exploit the modular operator and functional calculus to streamline the proof
following their underlying strategy.
We also extend this result to the maximum relative
entropy with respect to a fixed reference state which has important implications for studying
the superadditivity of the capacity of a quantum channel to transmit classical information.
\end{abstract}

      \pagebreak
 
\section{Introduction}  \label{sect:intro} 
It is now well established that the availability of entangled states for quantum systems implies that
many information theoretic quantities, such as the capacity of a quantum channel to transmit
 information, need not be additive under tensor products.  In this paper we consider
 the minimal output entropy   $S_{\min}(\Phi) $ associated with a quantum channel $\Phi$ which acts on 
 $M_d$ the set of $d \times d$ matrices over ${\bf C}$.
 A quantum state is described by a density matrix $\rho$, i.e., a positive semi-definite 
 matrix with $\tr \rho = 1$.
The quantum entropy of  the state is $ S(\rho) \equiv - \tr \rho \log \rho $. Then
\be    \label{Smin} 
  S_{\min}(\Phi) \equiv \inf \{ S[\Phi(\rho) ] :  \rho \geq 0, \tr \rho = 1 \} .
\ee 
In 2009, Hastings \cite{H} showed that  $S_{\min}(\Phi) $ is not additive, i.e.,  there are quantum channels $\Phi$
for which $  S_{\min}(\Phi \ot \Phi) < 2   S_{\min}(\Phi)$.  (See  \cite[Section 8.2.3] {AS} and endnotes to 
Chapter~8  of \cite{AS} for a discussion of later developments.)

Nevertheless,  Gour and Friedland \cite{GF} proved that the minimal output entropy of a quantum channel is locally additive. 
$S[\Phi(\rho)] $ has a local minimum at $\rho_c$ if there is a neighborhood $\cN(\rho_c)$ such that
$S[\Phi(\rho)] \geq S[\Phi(\rho_c)] $ for all $\rho \in \cN(\rho_c)$.  If the inequality is strict for all $\rho \neq \rho_c $
the minimum is non-degenerate.
Because the entropy $S(\rho) $ is a concave function and the set of density matrices is compact and convex, any minimum,
local or global, must be achieved at an extreme point so that $\rho_c = \proj{\psi} $ is a pure state.

The main result of \cite{GF}  can be stated as follows.  If a pair of quantum channels $\Phi_A, \Phi_C$ have   non-degenerate local minima of
$S[\Phi_A (\rho_A)]$  and $S[\Phi_C (\rho_C)]$  at $\rho_A = \proj{\psi_A} $ and
  $\rho_C = \proj{\psi_C} $ respectively, 
 then $ S \big[ (\Phi_A \ot \Phi_C)(\rho_{AC} ) \big] $ has a local minimum at $\rho_{AC} = \proj{ \psi_A \ot \psi_C} $.
    In this manuscript, we present this result with a number of simplifications to their proof.
 We also extend it to the maximum output for relative entropy with respect to a
 fixed reference state.
 
    In Section~\ref{sect:matrix} we  follow   \cite{GF} in using the Stinespring 
 representation to  reformulate the problem in terms 
 of the entanglement entropy of  a bipartite state  and then as the entropy $S(XX^\dag) $ when
 $X$  is a  $d_B \times d_E $ matrix restricted to a subspace of such matrices.   
 After doing this, we restate the main result as 
 Theorem~\ref{thm:main}.
 
 However, we use a completely different approach to reduce the general
 case to that of $X$  a square positive definite matrix.  This  reduction  
 is based on  the the fact that  $S(X^\dag X) = S(XX^\dag)$ and a simple majorization argument
 as described in Section~\ref{sect:nonsing}.  The further reduction to a positive definite matrix is given in  
Section~\ref{sect:polar} using the polar decomposition theorem; this section also presents
the framework needed later in Section~\ref{sect:nonsing}.  
  
When $d_B = d_E$  and $X$ is non-singular, we follow the underlying strategy of \cite{GF} in Sections~\ref{sect:deriv} to \ref{sect:proof}. 
 However, we make significant simplifications to their arguments.  In  Section~\ref{sect:deriv} we use 
  the integral representation of the $\log$ to easily obtain the relevant derivatives.
  We then  rewrite these derivatives using the elegant modular
 operator theory of Tomita-Takesaki as   advocated by D. Petz \cite{Petz}. 
 This approach avoids the need to write matrix elements explicitly.    
It also allows us to associate each critical point $X_c$ with an operator $H(X_c)$ which acts
 on a particular self-adjoint subspace of $M_d \op M_d $.  The condition that this
 critical point corresponds to a non-degenerate local minimum of $S(X X^\dag) $  is that this
 operator is positive definite on that subspace.
        We complete the proof in Section~\ref{sect:proof} with a cleaner presentation 
    using the relative modular operator. 
   
   In Section~\ref{sect:relent},  we extend our results in a straightforward way to 
 the local maxima of the relative entropy  $H[\Phi(\rho),\Phi(\omega)]$ 
 with $\omega$ fixed and discuss the implications for studying superadditivity 
 of the capacity of a quantum channel to transmit classical information.
 
 In Appendix~\ref{app:phi} we give two proofs of an elementary key inequality from \cite{GF}.
    In   Appendix~\ref{app:deriv}, we use an integral representation of $\log \rho(t)$ to find useful formulas 
    for its first derivative.   
 
 \section{Reformulations}  \label{sect:reform}

    \subsection{Matrix reformulation}   \label{sect:matrix}

Recall that a quantum channel $\Phi : M_{d_A} \mapsto M_{d_B} $ is a completely positive trace-preserving map 
which can be represented using an auxiliary space ${\bf C}_{d_E} $ of dim $d_E \leq d_A d_B$  via the Stinespring
representation in the form
 \be \label{stine}
  \Phi(\rho) = \trp_E \, K \rho K^\dag , 
 \ee where  $K: {\bf C}_{d_A} \mapsto   {\bf C}_{d_B} \ot {\bf C}_{d_E} $ satisfies  $K^\dag K = I_A $.
Since the minimal output entropy is the optimization of a concave function over a convex set, it suffices to consider pure 
inputs  $\proj{\psi}  $.   Then  $K | \psi \ket  \in {\bf C}_{d_B} \ot {\bf C}_{d_E} $.
  If we interpret $u, v $ as column vectors then  $ \Omega | u \ot v  \ket = u v^T $  extended by linearity  takes vectors in 
${\bf C}_{d_B} \ot {\bf C}_{d_E} $ to $d_B \times d_E $ matrices.
 If $|\psi \ket \in  {\bf C}_{d_A} $ and $X = \Omega\big( K |  \psi \ket  \big)$ then  $X :  {\bf C}_{d_E}  \mapsto  {\bf C}_{d_B} $ 
 so that $X$ is a $d_B \times d_E $ matrix and
 \be  \label{bipart}
 X X^\dag =    \Phi(\proj{\psi}) = \trp_E \, K \proj{\psi} K^\dag  \,.
\ee
 The entropy $S\big( X X^\dag\big) $
is the entanglement entropy of the bipartite state $K | \psi \ket $.  Thus the problem of finding
the minimal output entropy of a quantum channel  is equivalent to finding
the minimal entanglement entropy of
a  bipartite state   $| \psi_{BE} \ket  $ in the subspace $KK^\dag  ({\bf C}_{d_B} \ot {\bf C}_{d_E} )$  which is the range of $K$.
Moreover, if  $\cK \equiv  \Omega \big[ K K^\dag  ({\bf C}_{d_B} \ot {\bf C}_{d_E} ) \big]$ 
is the corresponding  subspace of $d_B \times d_E $ matrices  this is
equivalent to finding $\inf_{ X \in \cK^o}   S(X X^\dag) $  where $\cK^o = \{ X \in \cK : \tr X X^\dag = 1 \}$ is
 the unit sphere in $\cK$.  We have thus proved the following
\begin{thm}   \label{thm:reform}
Let $\Phi : M_{d_A} \mapsto M_{d_B} $ be a quantum channel and $\cK$ the subspace of $d_B \times d_E$ matrices
defined via the Stinespring representation as above.  Then the entropy $S[\Phi(\rho)]$  has a local minimum at 
$\rho = \proj{\psi} $ if and only if  $ S(X X^\dag)  $ with $X \in \cK^o$  has a local minimum at $ X = \Omega K | \psi \ket $.
\end{thm}

  In this framework $X^T \ovb{X}  = \Phi^C(\proj{\psi} ) $ is the output of the complementary channel.
  Since $   X X^\dag $ and $ X^\dag X$ have the same non-zero eigenvalues, $S(X X^\dag ) = S(X^\dag X) = S(X^T \ovb{X}) $.  
 Note that  $X^\dag  X =   \trp_B  \ovb{K} \proj{\ovb{\psi}} K^T$ is an output of the complementary channel associated with  $\ovb{K} $.

It follows from Theorem~\ref{thm:reform}  that the local additivity of the minimal output entropy of a pair of 
quantum channels when both minima are non-degenerate is equivalent to the following  
\begin{thm}  \label{thm:main}
 If $  S( X  X^\dag)\big|_{X \in   \cK^o_B }$ and   $ S( X X^\dag)\big|_{ X \in  \cK^o_C }$ have non-degenerate local minima 
 at $X_B$   and $X_C$ respectively, then
$S( X  X^\dag)\big|_{X \in   (\cK_B  \ot \cK_C)^o}$   has a non-degenerate local minimum at $ X_B   \ot X_C $.
\end{thm}
This result can be extended to some cases of degenerate local minima as discussed at the end of Section~\ref{sect:proof}.
In Sections~\ref{sect:deriv}  to \ref{sect:proof}, we prove Theorem~\ref{thm:main}   under the additional assumption 
that  $X_B$ and  $X_C$ are square positive definite matrices.    
In the next subsection we carefully describe the framework needed to consider the general case.
In Section~\ref{sect:nonsing}   we complete the proof by showing that the general case can be reduced to that of a 
square non-singular matrix.

\subsection{Reduction to square non-singular and positive definite forms}  \label{sect:polar}

Whether or not $ S(X X^\dag) \big|_{X \in \cK^o}$ has a local minimum at  $X_c $ is determined
by directional derivatives associated with perturbations of the form
 $X(t) \equiv \sqrt{1-t^2} X_c + t Y$ with  $X_c \in \cK^o, Y \in \cK^o $ and $\tr X_c  Y^\dag = 0 $.
Then $ S(X X^\dag ) \big|_{X \in \cK^o}$ has a local minimum at $X_c$
 if and only if  $g(t) = S[X(t) X(t)^\dag ] $
has a local minimum at $t = 0$ for all $Y \in \cK^o \cap X^\perp $.


 For $X_c \in \cK^o$, let  $P_B$ be the projection onto $(\ker X_c X_c^\dag)^\perp$ and 
  $ P_E$  the projection onto $(\ker X_c^\dag X_c)^\perp$ so that $X_c = P_B X_c P_E $.
  The following theorem, which is proved in Section~\ref{sect:nonsing}, implies  
   that to determine whether or not $ S(X X^\dag ) \big|_{X \in \cK^o}$  has a
    local minimum at $X_c$ it suffices to consider directional derivatives  $X(t) \equiv \sqrt{1-t^2} X_c + t Y$ 
    with $Y \in (P_B \cK P_E)^o$.

    \begin{thm}  \label{thm:reduce}  Let $X_c = P_B X_c P_E$ as above.  Then $ S(X X^\dag ) \big|_{X \in \cK^o}$  
    has a local minimum at $X_c$ if and only if $ S(X X^\dag ) \big|_{X \in  (P_B \cK P_E)^o}$  has a local minimum
    at $X_c$.
    \end{thm}

   Since $\tr P_B = \tr P_E \equiv d $,  this essentially reduces the problem to
$d \times d$ matrices on the restriction to  $P_B \cK P_E$ with $X$ having full
rank $d$.  Using a variant of the polar decomposition, we can   write
$X = \sqrt{X X^\dag}\,  V $ with $V \in M_d$ unitary.
    Since $V V^\dag = I_d$, we can replace  
    \be
    \sqrt{1-t^2} X + t Y =  (\sqrt{1-t^2} \sqrt{XX^\dag}  + t Y V^\dag ) V 
\ee
by $(\sqrt{1-t^2} \sqrt{XX^\dag}  + t Y V^\dag ) $.  Thus, we can assume that $X = \sqrt{X X^\dag}$ is positive
definite by changing the perturbation from  $Y$  to  $Y V^\dag$.

In terms of our original formulation, we can identify $V$  with a partial isometry  of rank $d$ with
$V V^\dag = P_B $ and $ V^\dag V = P_E $.  Then $Y \in P_B \cK P_E = P_B \cK V^\dag V$ 
if and only if  $Y V^\dag \in P_B \cK V^\dag $.    Note that  elements of $P_B \cK V^\dag $ 
map ${\bf C}_{d_B} \mapsto {\bf C}_{d_B} $ and  $\cK_B  \equiv P_B \cK V^\dag $  can be
identified with a subspace of $M_{d_B}$.  
Thus,  we can replace $X \in \cK$ by the positive definite matrix 
$X = \sqrt{X X^\dag} \in \cK_B $ by considering perturbations $Y \in \cK_B = P_B \cK V^\dag $.

\begin{remark}  \label{cor:posdef}  Thus 
$  S( X  X^\dag)\big|_{X \in   \cK^o }$ has a local minimum at  $X_c = \sqrt{X_c X_c^\dag } \, V  \in \cK^o$ if and only if  
  
   a) $ S( X  X^\dag)\big|_{X \in   \cK^o_B }$   has a local minimum at   $ | X_c^\dag | \equiv \sqrt{X_c X_c^\dag}  \in  \cK^o_B $  or, equivalently,
   
   b)  $ g(t) =  S[  X(t) X(t)^\dag ]$ with $X(t) =  \sqrt{1- t^2}  \sqrt{X_c X_c^{\dag} }+t Y  $    has a local minimum at 
      $t = 0$ for all $Y \in \cK^o_B \cap X^\perp$.
\end{remark} 

Thus, we can assume that $X$ is positive definite  and
consider $X(t) =  \sqrt{1-t^2} X + t Y $ with $X, Y$ in the unit sphere $\cK^o$ of some fixed subspace of $M_d$
and $\tr X Y^\dag = 0 $.  

If we write $Y = W + i Z $ with $W= W^\dag$ and  $Z = Z^\dag$, then  our assumption that $X > 0 $ implies that if 
$\tr XY^\dag = 0 $ then $\tr X W = \tr XZ = \tr X Y = 0 $.  As observed after \eqref{gam0}, 
the condition  which determines the critical points, requires only  perturbations
$W = W^\dag $ in  $ X^\perp$.  
Whether or not these critical points correspond to  non-degenerate local maxima 
requires only perturbations $W$ and $i Z $ with $W = W^\dag \in X^\perp$ and $Z = Z^\dag \in X^\perp$
as can be seen from \eqref{secdf}.

  \section{Derivatives  of entropy expressions}   \label{sect:deriv}

\subsection{Derivatives of the entropy}   \label{sect:intrep}

Let $\rho(t) $  be a one-parameter family of density matrices  
which is  twice differentiable  and has full rank in a neighborhood of $t = 0 $.  
The derivatives of $S[\rho(t) ]= - \tr \rho(t) \log \rho(t) $  can be easily obtained from
 those for $\log \rho(t)$.  For completeness,
  these are derived using an integral representation  for $ \log \rho$  in Appendix~\ref{app:deriv}.
 
It is well known \cite{Petz} and shown in Appendix~\ref{app:deriv}  that
 \be  \label{logderiv}
     \frac{d~}{dt} \log \rho(t) \Big|_{t = t_1} &  = & 
       \int_0^\infty    \frac{ 1}{\rho(t_1 ) + u I } \rho^\prime(t_1)  \frac{ 1}{\rho(t_1 ) + u I }  du   \,.
     \ee

  Next, observe that the derivative of  $\log \rho(t) $ is always orthogonal to $\rho(t)$, i.e.,
  \be   \label{int1}
     \tr  \rho(t) \int_0^\infty 
             \frac{ 1}{\rho(t ) + u I } \rho^\prime(t) \frac{ 1}{\rho(t  ) + u I } du   
               & = &  \tr    \int_0^\infty   \rho^\prime(t)
             \frac{ 1}{\rho(t ) + u I } \rho(t)  \frac{ 1}{\rho(t  ) + u I }     du  \qquad   \nn \\
             & = & \tr      \rho^\prime(t) \rho(t) \rho(t)^{-1}   \nn \\
             & = &  \tr    \rho^\prime(t)  = 0   
  \ee 
  where we used the cyclicity of the trace.

It follows immediately from \eqref{int1} that
\be  \label{deriv1}
      \frac{d~}{dt} S[\rho(t)  ] = - \tr \rho^\prime(t) \log \rho(t) - \tr \rho(t)  \frac{d~}{dt} \log \rho(t)  = - \tr \rho^\prime(t) \log \rho(t) 
      \ee
      \be  \label{deriv2}
         \frac{d^2~}{dt^2} S[\rho(t)  ]  & = &   - \tr \rho^{\prime \prime} (t) \log \rho(t) - \tr  \rho^\prime(t)  
             \int_0^\infty    \frac{ 1}{\rho(t ) + u I }   \rho^\prime(t) \frac{ 1}{\rho(t  ) + u I }   \,   du  \, .
             \ee
  
\subsection{Derivatives in matrix form}   \label{sect:matderiv}
  We now consider   $X, Y \in \cK^o $ the unit sphere on a subspace of $M_{d_B} $   as described in Section~\ref{sect:polar}
   with $ \tr X Y^\dag = 0$ and
define $X(t) \equiv \sqrt{1-t^2} X + t Y$, so that   $\tr X(t) X(t)^\dag = 1 $.  We also assume that  $X X^\dag$ is non-singular
so that we can apply \eqref{deriv1} and \eqref{deriv2}  to
          \be  \label{perturb}
\rho(t) 
& = &     (\sqrt{1 - t^2} \, X + t Y)(\sqrt{1 - t^2} \,  X+t Y)^\dag  \nn  \\
 & = &   (1- t^2) X X^\dag + t \sqrt{1 - t^2} (  Y X^\dag  +   X Y^\dag ) + t^2 Y Y^\dag  \qquad 
  \ee
to obtain
 \be  \label{D1}
  D_1[X,Y] \equiv   \frac{d~}{dt} S[\rho(t)  ] \bigg|_{t = 0} & = &  - \tr ( Y X^\dag   +   X Y^\dag ) \log X X^\dag= - \tr \Gamma_1 \log X X^\dag 
 \ee 
  \be   \label{D2}
  D_2[X,Y] \equiv   \frac{d^2~}{dt^2} S[\rho(t)  ] \bigg|_{t = 0} = - 2 \,  \tr \Gamma_2 \log X X^\dag-  
 \tr  \Gamma_1     \int_0^\infty   \frac{ 1}{X X^\dag + u I }  \Gamma_1  \frac{ 1}{X X^\dag + u I }   \,   du  
 \ee      
  where        $\Gamma_1 =   Y X^\dag   +   X Y^\dag  = \rho^\prime(0) $  and      
  $\Gamma_2 =  Y Y^\dag- X X^\dag = \half \rho^{\prime \prime}(0)$.  
 
 With the additional assumption that $X > 0 $ and $X, Y \in \cK_B^o$ as discussed above
 Remark~\ref{cor:posdef}, it is convenient to write
  \be  \label{D2RQ} 
   D_2[X,Y] = R[X,Y] - Q[X,Y]
  \ee
  where
  \be   \label{Rdef}
     R[X,Y] & \equiv &  - 2 \,  \tr (Y Y^\dag \log X^2) - 2 \, S(X^2)\\ \nn ~~ \\   \label{Qdef} 
   Q[X,Y] & = &  \tr  (Y X  +   X Y^\dag)   \int_0^\infty   \frac{ 1}{X^2 + u I }  (Y X +   X Y^\dag)  \frac{ 1}{X^2 + u I }  \, du  ~. 
  \ee
   Observe that $- 2 \,  \tr (Y Y^\dag \log X^2)$ is the only positive term in the second derivative.
Therefore,  if $S(X^2) $ is a local minimum, this term must dominate.
 
 We can easily extend all of the expressions above to situations when $\tr Y Y^\dag = \lambda^{-2} \neq 1$ by
 observing that $D_1[X, \lambda Y ] =   \lambda  D_1[X,Y] $ and will occasionally find it useful to do so.

          \subsection{Modular operator expression  for second derivative}  \label{sect:D2mod}
        For simplicity, we now  assume that $X = X^\dag > 0 $ is positive definite as discussed above
        Remark~\ref{cor:posdef}.
      
      We can rewrite   \eqref{Qdef}  in  an elegant and useful way using the modular operator
      formalism.  
           Let    $L_A(W)  = AW $ and $ R_B(W) = WB $ denote the operations of left and right multiplication  
           on $M_d$.  Since $L_A R_B(W) = AWB = R_B L_A(W)$, the operators
 $L_A, R_B$ commute  even when $AB \neq BA$.  
 When $A, B$ are Hermitian, positive semi-definite, or positive definite,
 then $L_A $ and $R_B$ are correspondingly Hermitian, positive semi-definite, or positive definite with respect to the
 Hilbert Schmidt inner product.  Moreover $f(L_A) = L_{f(A)}$ etc.  
 
     Now write $  Y   = W + i Z$ with\footnote{If we had not assumed that  $X> 0$, we could obtain an equivalent  result by writing
           $X = \sqrt{X X^\dag} \, V $  and letting $V Y^\dag  = W - i Z$ with 
           $W = \half(   V  Y^\dag +  Y V^\dag  )  $ and $Z =  i \half ( V  Y^\dag -  Y V^\dag  )    $.}
             $W = \half(      Y^\dag +  Y    ) = W^\dag $ and $Z =  i \half (    Y^\dag -  Y   ) = Z^\dag  $.
      Then
           \be  \label{gam0}
             \Gamma_1  & = &   (W + i Z)  X  +   X \, (W - i Z)  \nn \\
              & = &       (L_X+ R_X )( W )- i (L_X - R_X) (Z)   \\
              & =& \nn  M_+ (W )-  i M_- (Z )  \qquad   \hbox{with} \qquad M_\pm  = L_{ X} \pm  R_{ X} \, .
\ee
	It follows from \eqref{D1} that  $D_1[X,Y] =   D_1[X,W] = -2 \,  \tr W X \log X^2$ when $X > 0 $.

Next, observe that 
\be  \label{Rcross}
    R[X,Y] =  - 2 \tr (W^2 + Z^2 ) \log X^2 - 2 S(X^2)  + 2 i  \, \tr ( W Z - ZW ) \log X^2   .
\ee
To treat $Q[X,Y] $ we will use  the following well-known formula  \cite{Petz}  \cite[Appendix~\ref{app:phi}]{HKPR} 
\be   \label{BKM}
         \frac{\log L_\rho - \log R_\rho }{L_\rho - R_\rho}(\Gamma) = \frac{ \log \Delta_\rho}{L_\rho - R_\rho} (\Gamma) 
                           = \int_0^\infty \frac{1}{\rho+ tI}\,\Gamma\,\frac{1}{\rho+ tI}\,dt   
 \ee
where  $\Delta_\rho = L_\rho R_\rho^{-1} $ is the modular operator.   
 This follows from \eqref{preBKM} in Appendix~\ref{app:deriv} and holds whenever $\rho > 0 $ is positive definite.
Thus,   \eqref{Qdef} can be written as 
 \be   \label{D2b}
   Q[X,Y]  =      \tr  \Gamma_1    \frac{\log \Delta_{X^2} } {L_{X^2 } - R_{X^2 } } ( \Gamma_1)  = \tr \Gamma_1   \frac{\log \Delta_{X^2}}{M_+ M_{-}}  ( \Gamma_1 )   
    \ee
and the fact that  $M_\pm$ and $\Delta_{X^2} $ all commute implies that  \eqref{D2b}  can be written as
\begin{align}  \label{full}
 Q[X,Y]  & = 
        \tr  \big[ M_+ (W )-  i M_- (Z )   \big] \frac{\log \Delta_{X^2}  }{L_{X^2 } - R_{X^2}   }  
               \big[ M_+ (W )-  i M_- (Z )  \big]  \\ \nn
               & =      \tr M_+ (W) \frac{ \log \Delta_{X^2}}{M_+ M_{-}}  M_+ (W) -  \tr M_-(Z )\frac{ \log \Delta_{X^2}}{M_+ M_{-}}  M_- (Z)   \\
  & \qquad    - i  \,  \tr    M_+  (W) \frac{\log  \Delta_{X^2}}{M_+ M_{-}}  M_- (Z)    
 -i  \,  \tr  M_- (Z) \frac{\log  \Delta_{X^2}}{M_+ M_{-}} M_+ (W)  \, . \nn  \end{align}  
Since the cyclicity of the trace implies  $ \tr M_{\pm}(A) B = \pm  \tr A M_\pm(B)  $,  this gives
\begin{align}      \wtd{Q} [X,W+iZ]  &  \equiv Q[X,Y]   - 2 i \,  \tr (  WZ - ZW) \log X^2   \label{Qcross} \\  
  & =      \tr   W \,  M_+  \frac{\log \Delta_{X^2}}{M_+ M_{-}}  M_+(W)
    +  \tr Z \, M_-   \frac{ \log \Delta_{X^2}}{M_+ M_{-}}  M_-(Z)   \label{decoup}   \\ \nn
              & =         \tr   W \,  \frac{M_+}{  M_{-}}   \log \Delta_{X^2}
  W +  \tr Z \,  \frac{M_ {-} }{M_+  }  \log \Delta_{X^2}  Z  \\   
   & =      \tr   W \,  \frac{\Delta_{ X} + I }{\Delta_{ X} - I } \log  \Delta_{X^2}  \,  W   
       +   \tr Z \, \frac{\Delta_{ X} - I }{\Delta_{ X} + I } \log \Delta_{X^2}  \,   Z  \nn  \\   \label{QWZ}  
         & =      \tr W \, \phi( \Delta_{ X}) (W)  +  \tr  Z \, \phi(- \Delta_{ X}) (Z)  
                  \end{align}
where we used 
\be
  \frac{M_\pm}{M_\mp}=   \frac{ L_{ X} \pm
   R_{ X} }{ L_{ X}  \mp R_{ X } }
                  =   \frac{\Delta_{X} \pm 1}{  \Delta_{X} \mp 1  } 
 \ee 
and   define 
\be  \label{phidef} 
\phi(a) \equiv \frac{a+1}{a-1} \log a^2 ~.
\ee
 The properties of $\phi$, which plays a key role, are summarized in Appendix~\ref{app:phi},
where we also include two proofs of the following inequality of Gour and Friedland    \cite[Lemma 6]{GF}. 
 \be     \label{phikey}
  2  \phi(bc)     \leq  \phi(b) + \phi(-b) + \phi(c) + \phi(-c) .
\ee
Since $\phi(a) + \phi(-a) = \phi(a^2)$, this can be rewritten as
\be   \label{phikey2}
   2 \phi(bc) \leq \phi(b^2) + \phi(c^2) .
   \ee 
  
  Combining \eqref{Rcross}, \eqref{Qcross} and \eqref{QWZ} gives the following
    \begin{thm}  \label{thm:YiY}
    When $X > 0$ is positive definite, and $Y = W + i Z $
    \be   \label{secdf}
 D_2[X,Y]  &   \equiv  &  \frac{d^2~}{dt^2}  S \Big(  (\sqrt{1 - t^2} \, X + t Y)(\sqrt{1 - t^2} \,  X+t Y)^\dag   \Big) \bigg|_{t = 0 }  \nn  \\    
 & = &   - 2\, S(X^2)  
                 - 2 \, \tr(  W^2  + Z^2)  \log X^2 - \wtd{Q}[X,Y]  \\ \nn 
  & = &   - 2\, S(X^2)  
                 - 2 \, \tr(  W^2  + Z^2)  \log X^2
          -   \tr W \, \phi( \Delta_{ X}) (W)  -  \tr  Z \, \phi(- \Delta_{ X}) (Z)  \quad
              \ee
     with  $W, Z $ as defined above \eqref{gam0} and $\phi(a)$ in \eqref{phidef}.
       \end{thm}
\begin{remark} \label{D2alt}
We now make a number of useful observations with $Y = W + i Z $. 
\end{remark} 
 \begin{enumerate}[label=\alph{*})]
   
  
 \item $ 2 (W^2 + Z^2)  = Y Y^\dag + Y^\dag Y $.
 
\item   Since  $\phi(\pm  \Delta_{ X})$ is self-adjoint with respect to the Hilbert Schmidt inner product,
\begin{align*} 
   \tr W \, \phi(\pm  \Delta_{ X}) (W) +   \tr  Z \, \phi(\pm\Delta_{ X}) (Z)      
       & =     \tr (W + i Z)  \phi(\pm  \Delta_X) (W- iZ)  \\  
       &=   \tr Y  \phi(\pm  \Delta_X )( Y^\dag )  
         \end{align*}
      which implies  
    $    \tr W \, \phi( \Delta_X^2) (W) +   \tr  Z \, \phi( \Delta_X^2) (Z)     = \tr Y \phi( \Delta_X^2) (Y^\dag)  $.

\item
    We can also   write
  \be   \label{deltafin}
   \half   D_2[X, Y]  + \half D_2[X,iY]   & = &   - 2\, S(X^2)  - \tr (  Y Y^\dag  + Y^\dag Y)\log X^2 
       -   \half \tr Y [ \phi( \Delta_{ X}) + \phi(- \Delta_{ X}) ] (Y^\dag)   \nn \\
          & =& - 2\, S(X^2)  - \tr (  Y Y^\dag  + Y^\dag Y)\log X^2 -  \half \tr Y \phi(\Delta_X^2 ) Y^\dag \, .
 \ee

\item  Let $H_{\pm}(X) \equiv - 2 S( X^2) \, \id_d - 2 \, R_{\log X^2}  - \phi(\pm \Delta_X)   $ with $\id_d$  the
identity acting on $M_d$.  Then  when $\tr W^2 + \tr Z^2 = 1 $
\be    \label{d}
 D_2[X,W + i Z] =  \tr  (W, Z ) \pmx H_+(X)   & 0 \\ 0 & H_-(X)   \emx \pmx W \\ Z \emx ~.
 \ee 

\end{enumerate} 

As discussed in \cite{GF}, the minimal output entropy is not locally additive if we restrict to
vector space of matrices whose entries are real.   In  \eqref{d}, we replace $\cK$
by the real subspace of self-adjoint matrices in $\cK \op \cK $.
The complex structure is implicit in our use of $\phi(+\Delta_X) $ in the first block and  $\phi(-\Delta_X) $ in the second block
of  $H(X)  =  \pmx H_+(X)   & 0 \\ 0 & H_-(X)   \emx  $.
Then $D_2[X_c,Y] > 0 $ for all $Y \in X_C^\perp$ is equivalent to the strict positivity of
$H(X_c) $ on the subspace of self-adjoint matrices in $\cK \op \cK  \cap \{ (X_c,0) \cup (0, X_c ) \}^\perp  = X_c^\perp \op X_c^\perp$.  Let  $X_c > 0 $ and $D_1[X_c, W] = 0  $ 
for all $W = W^\dag \in \cK^o$ so that
$X_c$  is a critical point of $S(X X^\dag)$.   The strict positivity of $H(X_c) $ on the subspace of self-adjoint matrices in $X_c^\perp \op X_c^\perp$ 
means that there is a $\nu > 0 $ such that $H(X_c) > \nu I_{2d-2} $ 
on  this subspace which implies
that  $D_2[X_c, Y] > \nu $ for all $Y \in \cK^o \cap X_c^\perp$.  Thus, with $X(t) = \sqrt{1-t^2} X_c + t Y $ it follows from Taylor's theorem  with remainder that
\be   \label{taylor} 
 S(X(t) X(t)^\dag ) >  S(X_c^2) + \half \nu \, t^2 - \tfrac{1}{6} R \,  |t|^3  \qquad  \forall ~~Y \in \cK^o \cap X_c^\perp   \, .
   \ee
if  $  | \frac{d^3~}{dt^3} S(X(t) X(t)^\dag ) | < R $.    We show in Appendix~\ref{app:D3bound}
that  there is an  $R > 0 $ and    $\tau \in  (0,1) $ such that  
this holds for all $t \in (-\tau,\tau ) $.   Thus, $ S(X(t) X(t)^\dag ) >  S(X_c^2)  $  when $0 < |t| < \min \{ \tau,  \tfrac{ 3 \nu}{R} \} $.
Since $S( X(0) X(0)^\dag ) =  S(X_c^2) $, this implies that $ S(X X^\dag) $ has a non-degenerate local minimum at $X_c$.

\section{First derivative essentially additive}  \label{sect:first}
We now assume that
 $  S( X^2 ) \big|_{X \in   \cK^o_B }$ and   $ S( X^2)\big|_{ X \in  \cK^o_C }$ have local minima 
 at $X_B > 0 $   and $X_C > 0 $ so that
  $D_1[X_B, Y_B] = D_1[X_C, Y_C]  = 0 $ for all $Y_B \in  \cK_B^o \cap X_B^\perp$  and $Y_C \in  \cK_C^o \cap X_C^\perp $.
   Then for all $Y_{BC} \in      (\cK_B \ot \cK_C)^o \cap (X_B \ot X_c)^\perp $
     \be
  D_1[X_B \ot X_C, Y_{BC} ]  
    & = &  - \trp_{BC} \big[ ( X_B \ot X_C)  Y_{BC}^\dag +  Y_{BC} (X_B \ot X_C)    \big] \log (X_B^2 \ot X_C^2)  \nn  \\
       & = & -  \trp_B \, \big(  X_B   \wtd{Y}_B^\dag + \wtd{Y}_B X_B  \big) \log X_B^2
          -   \trp_C \,   \big( X_C \wtd{Y}_C^\dag + \wtd{Y}_C X_C  \big) \log X_C^2   \nn \\
          & = & D_1[X_B, \wtd{Y}_B] + D_1[X_C, \wtd{Y}_C]  = 0 + 0 ~ = ~  0
  \ee
where $ \wtd{Y}_B = \trp_C \, X_C Y_{BC} $ and $\wtd{Y}_C = \trp_B \, X_B Y_{BC} $ and, e.g., 
$ \trp_B X_B \wtd{Y}_B^\dag  = \trp_{BC} \, (X_B \ot X_C) Y_{BC}^\dag = 0 $.

Note that we did not find  $D_1[X_B \ot X_C, Y_{BC} ]  =  D_1[X_B, Y_B] + D_1[X_C, Y_C] $.
Instead we used the fact that the directional derivatives with respect to all allowed perturbations are zero.
We will use a similar strategy to treat the second derivative.


 \section{Second derivative increases}  \label{sect:proof}

\subsection{Key inequality of Gour-Friedland}

    To simplify the notation  we  write $\Delta_B$ for $\Delta_{X_B}$ etc.
    and use $\id $ to denote the identity operator acting on $d \times d $ matrices.     The key result needed to treat $\wtd{Q}[X,Y] $  is
        \be   \label{DeltaKey}
   2   \phi(\pm \Delta_B \ot \Delta_C) \leq  \phi(\Delta_B^2 ) \ot \id_C +    \id_B \ot  \phi(\Delta_C^2) \, .
             \ee
   This follows immediately from  \eqref{phikey2}  and the  spectral theorem applied  
   to $\Delta_B$ and  $\Delta_C$ 
   
This  inequality is a substitute for additivity.  It implies subadditivity of $\wtd{Q}[X,Y] $ 
which leads to superadditivity of  $D_2[X,Y]$.  The presence  of either $\id_B $ or $\id_C $ 
also allows us to eliminate  many cross terms in $\wtd{Q}[X,Y] $.

In what follows we will make frequent use of the fact that for $X > 0 $ both $\Delta_X $
and $\phi(\Delta_X) $ are self-adjoint with respect to the Hilbert-Schmidt inner product.
Moreover,  $\Delta_X(X) = X$  implies that 
$\phi(\Delta_B \ot \Delta_C)( X_B \ot Y_C) = X_B \ot \phi(\Delta_C)(Y_C)$.  We will frequently combine
this with self-adjointness  to conclude that when $W = W^\dag$ 
\be  \label{crosskey} 
  \trp_{BC} \,  (X_B \ot W_C )  \big[  \phi(\Delta_B \ot \Delta_C)( F_{BC} ) \big]^\dag = \trp_{BC}  \, X_B \ot  \phi(\Delta_C)(W_C) \, F_{BC}^\dag \, .
\ee
As previously observed, since $X $ is self-adjoint  $\tr  X Y^\dag = 0  \Leftrightarrow \tr XY = 0$ and similarly for $W, Z$.

 We will also use the following result.   When  $D \in \cA \ot \cB$ this is essentially the singular value 
 decomposition of the operator that maps $ A \in \cA $ to $ \trp_{\cal A}  \, A^\dag D \in \cB$.
\begin{remark}   \label{SVD}
Let $\cA $ and $\cB$ be vector spaces of matrices  and $ D  \in \cA \ot \cB $.
Then one can always write  $D = \sum_k \omega_k A_k \ot B_k $ with
$\omega_k \in {\bf C} , \, A_k \in \cA,  \, B_k \in \cB$ and $\tr A_j^\dag A_k = \tr B_j^\dag B_k = 0$ when $j \neq k$.
\end{remark}

\subsection{Proof}    \label{sect:pseudo}

We  assume that $  S( X  X^\dag)\big|_{X \in   \cK^o_B }$ and   $ S( X X^\dag)\big|_{ X \in  \cK^o_C }$ have 
non-degenerate local minima   at $X_B$   and $X_C$ respectively.

First, observe that for $X_B  \in \cK_B^o $   and $X_C \in \cK_C^o $, 
\be
     ( X_B  \ot X_C) ^\perp = \spn  \{ ( X_B  \ot  X_C^\perp)   \cup ( X_B^\perp \ot X_C ) \cup  (X_B^\perp \ot X_C^\perp) \}
    \ee
where it is understood that $X_B^\perp$ and $X_C^\perp$ are subspaces of $\cK_B  $ and $\cK_C  $ respectively.
Remark~\ref{SVD} implies that any matrix  in  $ ( X_B^\perp \ot X_C^\perp )^o$ can be written in the form
\be   \label{BCperp}
     T_{BC} = \sum_{j > 0 } \xi_j Y_B^j \ot Y_C^j      \hbox{~~ with ~~} \tr Y_B^j (Y_B^k )^\dag = \tr Y_C^j (Y_C^k )^\dag  = \delta_{jk} \, .
\ee
 Therefore, the most general perturbation we need to consider can be written in the form
 \be     \label{genpert}
      Y_{BC} =   u_1 X_B  \ot Y_C^0 + u_2 Y_B^0 \ot X_C +   \eta   T_{BC} 
     =   u_1 X_B  \ot Y_C^0 + u_2 Y_B^0 \ot X_C  + \eta  \sum_{j > 0 } \xi_j    Y_B^j \ot Y_C^j    
     \ee
 where  $  Y_B^0  \in (X_B^\perp)^o$ and $Y_C^0 \in (X_C^\perp)^o $.  
 Note, however, that $\tr Y_B^0  (Y_B^j )^\dag $  and  $\tr Y_B^0  (Y_B^j)^\dag $ 
 are  {\em not} zero in general.   Because  $Y_B^j \in X_B^\perp$ and $X > 0 $,  $\tr X_B (Y_B^j)^\dag = \tr X_B Y_B^j = 0 $ and 
 similarly for $Y_C^j $.  We assume that $|u_1|^2 +  |u_2|^2 + |\eta|^2  = 1$ and 
 $\sum_{j > 0}  |\xi_j|^2  = 1 $ so that 
  $\trp_{BC} Y_{BC}^\dag Y_{BC} = 1$.

 \begin{lemma}   With $Y_{BC} $ defined as in \eqref{genpert}
   \begin{align}   
     D_2[X_B \ot X_C, Y_{BC} ] & =   |u_1|^2  D_2 \big[X_B \ot X_C, X_B  \ot Y_C^0 \big] +
       |u_2|^2  D_2 \big[X_B \ot X_C, Y_B^0 \ot X_C \big]    \nn    \\  & \quad   
     + ~  |\eta|^2       D_2 \big[X_B \ot X_C,T_{BC}\big]    \label{nocross} \\
     & =  |u_1|^2 D_2[X_C,Y_C^0]  + |u_2|^2 D_2[X_B ,Y_B^0]   
          + |\eta|^2       D_2 \big[X_B \ot X_C, T_{BC}\big]   \label{D2red}  
 \end{align}
 \end{lemma}

\noindent{\bf Proof:}.   We first consider the
cross terms in $\wtd{Q}[X_B \ot X_C , Y_{BC} ] $.  Observe that 
\bee
W_{BC} = \half(Y_{BC} + Y_{BC}^\dag)  = u_1 X_B \ot W_C^0 + u_2 W_B^0 \ot X_C + \eta \half(T_{BC} + T_{BC}^\dag ) 
\eee
The cross-term with coefficient $u_1 \ovb{u}_2 $ is
\bee
    \trp_{BC} \, X_B \ot W_C^0  \phi(\Delta_B \ot \Delta_C) W_B^0 \ot X_C = \trp_{BC} \, X_B W_B^0 \  \ot \phi(\Delta_C)(W_C^0)  X_C = 0
\eee
where we used \eqref{crosskey} and  $ \tr_B X_B W_B^0  = 0 $.
Similarly,  we find
\be
      \trp_{BC}\, ( X_B \ot W_C^0) \big[ \phi(\Delta_B \ot \Delta_C) (Y_B^j \ot Y_C^j) \big]^\dag = 
            \trp_{BC}  X_B (Y_B^j )^\dag  \ot\phi(\Delta_C )(Y_C^0 ) (Y_C^j)^\dag  = 0            
 \ee
which is easily seen to imply that $ \trp_{BC}( X_B \ot  W_C^0) \phi(\Delta_B \ot \Delta_C)  W_{BC}^j  = 0 $.
Similar arguments hold with $Z_{BC} $.  Thus we conclude that all cross terms in 
$ \wtd{Q} (X_B \ot X_C , Y_{BC} )$ with
coefficient     $u_1 \ovb{u}_2 ,  u_1 \ovb{\eta} ,   u_2 \ovb{\eta} $ and their conjugates are zero.   
  
Next we consider cross terms 
 arising from $\tr (Y_{BC} Y_{BC}^\dag + Y_{BC}^\dag Y_{BC} ) \log (X_B^2 \ot X_C^2) $.
 Those  with coefficient  $u_1 \ovb{u}_2 $
 are easily seen to be zero.   Those  involving  $u_k \ovb{\eta} $ are trickier.
Let  
\bee    \Upsilon_1 & = &        u_1 \ovb{\eta}  ( X_B \ot Y_C^0 ) (\xi_j Y_B^j \ot Y_C^j)^\dag +   \ovb{u}_1 \eta  (\xi_j Y_B^j \ot Y_C^j) (X_B \ot Y_C^0)^\dag  \\
     \Upsilon_2  & = &   u_1 \ovb{\eta}   (\xi_j Y_B^j \ot Y_C^j)^\dag ( X_B \ot Y_C^0 ) +   \ovb{u}_1 \eta  (X_B \ot Y_C^0)^\dag  ( \xi_jY_B^j \ot Y_C^j) \, .
     \eee
     Then we find
\bsq \be 
  \lefteqn{  \trp_{BC}   \Upsilon_1   \log ( X_B^2 \ot X_C^2)  }\qquad \quad  \nn   \\
       & = &    \trp_{BC} \big[  u_1  \ovb{\eta }( X_B \ot Y_C^0 ) \,(\xi_j Y_B^j \ot Y_C^j)^\dag + 
                \ovb{u}_1 \eta  (\xi_j Y_B^j \ot Y_C^j)  (X_B \ot Y_C^0)^\dag  \big]  \log X_B^2  \label{crossYa} \qquad  \\ \label{crossYb}
       & ~ &  +   \trp_{C} \big[ ( \trp_B \, X_B Y_B^j) u_1  \ovb{\eta} Y_C^0 ) ( Y_C^j)^\dag + (\trp_B Y_B^j X_B )   \ovb{u}_1 \eta   Y_C^j ( Y_C^0) ^\dag \big]    \log X_C^2 .
     \ee \esq
 The last term \eqref{crossYb} is zero because $\trp_B X_B Y_B^j = \trp_B X_B (Y_B^j)^\dag = 0 $. 
  If we let $\omega = \trp_C  \,  \ovb{\xi}_j Y_C^0 (Y_C^j)^\dag $. and $\wtd{Y}_B  =   \ovb{u}_1\eta \,  \ovb{\omega} Y_B^j $,
  the  first term \eqref{crossYa} can be rewritten as 
     $$   \trp_B\ \big(  X_B  u_1   \ovb{\eta \, } \omega (Y_B^j)^\dag+    \ovb{u}_1 \eta \,  \ovb{\omega} Y_B^j X_B \big) \log X_B^2 = D_1[ X_B , \wtd{ Y}_B] = 0$$
   since $X_B$ is a critical point. 
   Because, e.g.,  $\trp_C Y_C^0 (Y_C^j)^\dag = \trp_C   (Y_C^j)^\dag Y_C^0$ , one similarly finds 
   $\trp_{BC}  \Upsilon_2  \log ( X_B^2 \ot X_C^2)    = D_1[ X_B , \wtd{ Y}_B] = 0$ so that the
   total contribution from these cross-terms is
  $  \trp_{BC} (  \Upsilon_1  +    \Upsilon_2)   \log ( X_B^2 \ot X_C^2)   = 0$.        Note that
  the cross-terms with coefficient $u_1 \ovb{\eta} $ need not be zero.  They must be combined 
   with those with coefficient $\ovb{u}_1 \eta $ and, moreover,  $u_1$ and $\eta$ are needed
   in the definition of $ \wtd{ Y}_B$.
   The corresponding cross terms involving  the  term
   with coefficient $u_2$   can be handled in the same way.  Thus, all  cross terms 
 arising from $\tr (Y_{BC} Y_{BC}^\dag + Y_{BC}^\dag Y_{BC} ) \log (X_B^2 \ot X_C^2) $ are zero.
 
 To complete the proof let $C(t) = \sqrt{1-t^2} X_C + t  Y_C^0$  and observe that the entropy associated with the first term is
$ S\big( X_B^2 \ot C(t)^\dag C(t) \big) = S(X_B^2) + S\big( C(t)^\dag C(t) \big) $ so that
$D_2[X_B \ot X_C , X_B \ot Y_C^0] = D_2[X_C , Y_C^0] $.  A similar result holds for the second term
which gives \eqref{D2red}.      \qed

 We now consider $ D_2[X_B \ot X_C , T_{BC}]$ and write $ T_{BC} = W_{BC} + i Z_{BC} $  as before.    Then   
 \eqref{DeltaKey} implies
 \be
   2 \,  \wtd{Q} ( \Delta_B \ot \Delta_C, T_{BC} ) & = &    2 \trp _{BC}\, W_{BC} \phi( \Delta_B \ot \Delta_C ) W_{BC}   +  2 \,  \trp _{BC} \, Z_{BC} \phi(- \Delta_B \ot \Delta_C ) Z_{BC}  \nn \\
        & \leq &     \trp _{BC} W_{BC} \big( \phi( \Delta_B ^2)  \ot \id_C \big) W_{BC}  +  \trp _{BC}  W_{BC}  \big( \id_B \ot \phi(\Delta_C^2 \big) W_{BC}   \nn \\
                 & ~ & +    \trp _{BC}\, Z_{BC} \big( \phi( \Delta_B ^2)  \ot \id_C \big) Z_{BC}  +  \trp _{BC}\, Z_{BC} \big(  \id_B \ot \phi(\Delta_C^2) \big ) Z_{BC}     \nn \\
                 & = &   \trp _{BC}\,  T_{BC} \big( \phi( \Delta_B ^2)  \ot \id_C \big)T_{BC}^\dag  +     \trp _{BC} \, T_{BC}\big(  \id_B \ot \phi(\Delta_C^2) \big)T_{BC}^\dag \nn \\
                 & = &  \sum_{j > 0} |\xi_j|^2 \Big(  \tr_B Y_B^j   \phi( \Delta_B ^2) ( Y_B^j)^\dag  +   \tr_C Y_C^j   \phi( \Delta_C ^2) ( Y_C^j)^\dag  \Big)  \label{Qred}
\ee               
   where the last two lines  follow  from Remark~6b and 
\bee  \trp _{BC}\,  T_{BC} \big( \phi( \Delta_B ^2) \ot \id_C \big) T_{BC}^\dag & = &
 \sum_{j> 0}   \sum_{k > 0}  \xi_j  \ovb{\xi}_k    \trp _{BC}   Y_B^j \ot Y_C^j \big( \phi( \Delta_B ^2)  \ot \id_C \big) (Y_B^k \ot Y_C^k )^\dag  \\
  & = &   \sum_{j> 0}    |\xi_j|^2 \tr_B Y_B^j   \phi( \Delta_B ^2) ( Y_B^j)^\dag      
  \eee
since $\tr_C Y_C^j (Y_C^k)^\dag    = \delta_{jk} $.  A similar argument holds for the term with      $\id_B \ot \ \phi(\Delta_C^2) $.     

We similarly find that    since   $ \log (X_B^2 \ot X_C^2) =  \log (X_B^2) \ot  I_C + I_B \ot \log (X_C^2 ) $
\be
\lefteqn{  \trp_{BC}   \big( Y_{BC} Y_{BC}^\dag + Y_{BC}^\dag Y_{BC} \big) \log (X_B^2 \ot X_C^2)   = }     \label{YYred} \\   \nn
    & ~ &       \sum_{j > 0} |\xi_j|^2 \trp_B  \big (Y_B^j (Y_B^j)^\dag + (Y_B^j)^\dag Y_B^j  \big) \log X_B^2                
            ~ + ~   \sum_{j > 0} |\xi_j|^2  \trp_C \big (Y_C^j (Y_C^j)^\dag + (Y_C^j)^\dag Y_C^j  \big) \log X_C^2   \quad            
\ee                                  
 Combining \eqref{Qred} and \eqref{YYred} and noting  $ S(X_B^2 \ot X_C^2)   = \sum_{j > 0 } |\xi_j|^2  \big[  S(X_B^2)   + S(X_C^2   ) \big]   $ we find
 \be
 \lefteqn{  D_2[X_B \ot X_C, T_{BC} ]  \geq }    \\     \nn 
   & + &   \sum_{j > 0 } |\xi_j|^2  \Big( -2 S(X_B^2)   + 
           \trp_B  \big[  Y_B^j (Y_B^j)^\dag + (Y_B^j)^\dag Y_B^j  \big] \log X_B^2  - \half  \trp_B Y_B^j   \phi( X_B ^2) ( Y_B^j)^\dag  \Big)  \\  \nn
             & + &   \sum_{j > 0 } |\xi_j|^2  \Big( -2 S(X_C^2)   +  
 \trp_C  \big[ Y_C^j (Y_C^j)^\dag + (Y_C^j)^\dag Y_C^j  \big] \log X_C^2  - \half  \trp_C Y_C^j   \phi( X_C ^2) ( Y_C^j)^\dag  \Big)  \\
           & = &  \half   \sum_{j > 0 } |\xi_j|^2  \big( D_2[X_B, Y_B^j]  +   D_2[X_B, i Y_B^j]  +  D_2[X_C, Y_C^j]  +  D_2[X_C, i Y_C^j]  \big)   \label{D2Tfin}
 \ee
 where we used \eqref{deltafin}.
 Combining \eqref{D2red} and{\ \eqref{D2Tfin} we find that
\be
\lefteqn{  D_2[X_B \ot X_C, Y_{BC} ]   \geq    |u_1|^2 D_2[X_C,Y_C^0]  + |u_2|^2 D_2[X_B ,Y_B^0]  }  \qquad   \nn  \\
         & ~ & +| \eta|^2   \half   \sum_{j > 0 } |\xi_j|^2  \big( D_2[X_B, Y_B^j]  +   D_2[X_B, i Y_B^j]  +  D_2[X_C, Y_C^j]  +  D_2[X_C, i Y_C^j]  \big) \geq 0 \quad 
 \ee
 with strict inequality if both minima are non-degenerate.
 
    This completes the proof of Theorem~\ref{thm:main}  when $X$ is square and nonsingular.
As observed at the end of  Section~III of \cite{GF}, this argument also shows that if one of 
the local minima is non-degenerate, and the other degenerate, then the product is a degenerate local minimum.

We emphasize again that we do not compare
$ D_2[X_B \ot X_C, Y_{BC} ] $ to  $ D_2[X_B , Y_B ] + D_2[X_C , Y_C ] $.  Instead
we use the fact that the second derivative is positive for arbitrary perturbations in $\cW_B$
and $\cW_C$ so that we can  bound the second derivative by a sum of  positive terms.

 \section{Reduction to square non-singular form}  \label{sect:nonsing}

 \subsection{$X X^\dag $ or $X^\dag X$ non-singular}  \label{reverse}

 {  As in Section~\ref{sect:polar}, $P_B$ and $P_E $ are the projections onto $(\ker X X^\dag)^\perp$ and 
$(\ker X^\dag X)^\perp$ respectively so that $X = P_B X P_E $.   
  We can write $X$ and $Y$ in block form as
  \bee
      \pmx  X_{11} & 0 \\ 0 & 0 \emx \qquad \qquad \pmx Y_{11} & Y_{12} \\ Y_{21} & Y_{22} \emx
  \eee      \nopagebreak
  where we identify  
 $   X_{11}  \simeq   P_B X P_E,     Y_{11}    \simeq   P_B Y P_E ,    Y_{12}   \simeq    P_B Y P_{E^\perp},  
       Y_{21}    \simeq P_{B^\perp} Y P_E ,        Y_{22}   \simeq  P_{B^\perp} Y P_{E^\perp} $. }    
  If $X X^\dag$ is non-singular so that $P_B = I $,  we can write $X = \pmx X_{11} & 0 \emx, Y = \pmx Y_{11} & Y_{12} \emx $
  and observe that $\Gamma_1 = X_{11} Y_{11}^\dag + Y_{11} X_{11} ^\dag $.  Thus, $Y_{12} $ has no effect on $D_1[X, Y] $ and  its only
  contribution to the second derivative is  $- \tr Y_{12} Y_{12}^\dag \log X X^\dag \geq 0 $.  Since non-zero $Y_{12} $
  can only increase the second derivative and, hence, the entropy, there is no loss of generality in assuming
  that $Y_{12} = 0 $, i.e.,  it suffices to consider $Y \in P_B \cW P_E $. 
  
   If $X^\dag X $ is non-singular, then we can exploit the fact that
  $S(X X^\dag) = S(X^\dag X) $ and use the same argument with
 $X^\dag = \pmx X_{11}^\dag & 0 \emx, Y^\dag= \pmx Y_{11}^\dag & Y_{21}^\dag \emx $   
to show that there is no loss of generality in assuming that $Y_{21} = 0 $.  Thus, it
again  suffices to consider $Y \in P_B \cW P_E $.  
 
 \subsection{General case}
 
 To deal with the general case, we assume $d_E \geq d_B$ and replace $X$ by 
 $\pmx  X_\eps & 0 \\ 0 & \eps F \emx $ with $F F^\dag = P_{B^\perp} ,
 F^\dag F = P_E^\perp$
 and   $X_\eps = (1- \eps^2 \tr P_{B^\perp})^{1/2}  \, X_{11} $. As discussed in
 Section~\ref{sect:polar}, there is no loss of
 generality in assuming that $X_{11}$ is positive definite.
   \bee
            X X^\dag = \pmx  X_\eps^2   & 0 \\ 0 & \eps^2 P_{B^\perp} \emx   \qquad  \qquad
           \log X X^\dag = \pmx  \log X_\eps ^2 & 0 \\ 0 &  P_{B^\perp} \log \eps^2 \emx \\  ~~\\
             \Gamma_1 = \pmx   X_\eps Y_{11}^\dag + Y_{11} X_\eps & X_\eps Y_{21}^\dag + \eps Y_{12} F^\dag \\
         Y_{21} X_\eps  + \eps F Y_{12}^\dag   & \eps ( F Y_{22}^\dag  + Y_{22} F^\dag ) \emx   \qquad \qquad  \quad
  \eee
 It is straightforward to see that as $\eps \raw 0 $
 \bee   
    D_1[X,Y] & = &  -  \tr   (  X_\eps Y_{11}^\dag + Y_{11} X_\eps)  \log X_\eps^2     
       +    \tr  F Y_{22}^\dag  + Y_{22} F^\dag ) \eps \log \eps^2  \\
    & \lraw &  \tr (X_{11} Y_{11}^\dag + Y_{11} X_{11} ) \log X_{11}^2 = D[X_{11}, Y_{11}]
    \eee
Thus, only $Y_{11} $ affects the first derivative.

 The treatment of the second derivative is more complex.  The terms \eqref{Rdef} and   \eqref{Qdef} become
  \be   \label{Reps}
  R[X,Y] & \equiv &  - 2  \,  \tr \big( Y_{11} Y_{11} ^\dag  +  Y_{12} Y_{12}^\dag \big)   \log  X_\eps^2
             - 2    \tr \big(Y_{21} Y_{21} ^\dag  +  Y_{22} Y_{22}^\dag \big) \log \eps^2  \nn    \\
    & ~ &  - 2 S(X_\eps^2)  +2  \eps^2 \log \eps^2 \,  \tr P_B^\perp 
 \ee
and
 \bsq \be
 \lefteqn{ Q[X,Y] = \tr \!  \int_0^\infty  \! \!   (X_\eps Y_{11}^\dag + Y_{11} X_\eps) \frac{P_B}{X_\eps^2+ u P_B } 
 (X_\eps Y_{11}^\dag + Y_{11} X_\eps^\dag) \frac{P_B}{X_\eps^2+ u P_B }  du } \qquad \qquad \quad \label{singa} \\
  & + &   2 \,  \tr \!  \int_0^\infty  \! \! ( X_\eps Y_{21}^\dag + \eps Y_{12} F^\dag ) \frac{P_{B^\perp}}{P_{B^\perp} (\eps^2 + u) }  
     (    Y_{21} X_\eps  + \eps F Y_{12}^\dag ) \frac{P_B}{X_\eps^2+ u P_B } du  \qquad  \qquad \label{singb}  \\
      & + &   \tr \!  \int_0^\infty  \! \! ( \eps ( F Y_{22}^\dag  + Y_{22} F^\dag ) \frac{P_{B^\perp}}{ \eps^2 + u }  
      \eps ( F Y_{22}^\dag  + Y_{22} F^\dag ) \frac{P_{B^\perp}} {\eps^2 + u } du  \label{singc} 
 \ee   \esq
 One  can omit $P_B$ and $P_{B^\perp} $ when they are sandwiched between matrices which
are invariant under $P_B$ and $P_{B^\perp}  $ respectively.   

The terms  in \eqref{Reps} and \eqref{singa} involving  $Y_{11} $ converge to $D_2[X_{11},Y_{11}]$ when $\eps \raw 0$.

To analyze \eqref{singb} observe that
  \be \label{singb3}
   \int_0^\infty   \frac{1}{ \eps^2 + u }   \frac{P_B }{X_\eps^2 + u P_B } \, du = \frac{P_B }{X_\eps^2 - \eps^2 } \big( \log X_\eps^2 - \log \eps^2  \big) ~.
 \ee
Therefore the terms with $\eps Y_{12} \raw 0$  as $\eps \raw 0$ so that
 the only contribution of  $Y_{12} $  to $D_2[X,Y] $ is $-\tr Y_{12} Y_{12}^\dag  \log X^2 $
which is positive.   
Although the terms with $Y_{21} $ in \eqref{Reps} and \eqref{singb} are not well-behaved, we can
exploit the fact that $S(X X^{\dag})  = S(X^{\dag}  X) $.   Since the exchange $X \leftrightarrow X^\dag $ gives
$Y_{12}  \leftrightarrow Y_{21}^\dag$ we find that, as above, 
the only contribution of $Y_{21} $ to $D_2[X,Y] $ is $-\tr Y_{21}^\dag Y_{21} \log X^2 $
which is positive.    Hence, there is  no loss of generality in assuming that  $Y_{12} $ and $Y_{21} $ 
are both zero.

To treat $Y_{22} $, observe that the eigenvalues of $    \pmx  X_{11}^2 & 0 \\ 0 & 0 \emx $ majorize
those of $  \pmx    \xi X_{11}^2   & 0 \\ 0 & \eps^2 Y_{22} Y_{22}^\dag \emx  $
 when $\eps$  is sufficiently small  and $\xi = \big( 1 - \eps^2 \tr Y_{22} Y_{22}^\dag \big)^{1/2} $.
 Thus perturbations in which  $Y_{22} $ is  the only non-zero block  always increase the entropy.  
 Since we showed above that we can assume that  $Y$ is block diagonal, this 
suffices to conclude that non-zero  $Y_{22}$  can only increase the entropy.  (We
could observe that the contribution of $Y_{22} $ to $Q[X,Y] $ above is completely decoupled 
from those of $Y_{12} $ and $Y_{21} $ so that there is no loss of generality in assuming  that $Y$ is block diagonal.)
  
 Thus, we have shown that if $D_2[X_{11},Y_{11} ] $ is positive for all $Y_{11} \in  P_B \cK P_E$, then
 $D_2[X,Y] $ is positive for all $Y \in \cW$.  This completes the proof of Theorem~\ref{thm:reduce}.

  \subsection{Refined treatment of \eqref{singb} and \eqref{singc} }

  Although not needed for the proof, we can say more about the divergent terms in $D_2[X,Y] $. 
  
 First, observe that using \eqref{singb3}  in \eqref{singb} with $Y_{12} = 0 $ and ignoring terms of order $\eps^2$ 
 implies that the contribution of $Y_{21} $ to  $Q[X,Y] $ is
   \be   \label{singintfinal}
 2 \, \tr  X_\eps   Y_{21}^\dag Y_{21} X_\eps(X_\eps^2 )^{-1} \big( \log X_\eps^2 - \log \eps^2  \big)  
   & = &     - 2 \, \tr    Y_{21}^\dag Y_{21}  \log \eps^2   +   2 \, \tr Y_{21}^\dag Y_{21} \log X_\eps^2 
       \ee
so that the net contribution of $Y_{21}$ to $D_2[X, Y]  =R[X,Y] - Q[X,Y] $ is  
\be
  - 2 \,  \tr    Y_{21}Y_{21}^\dag  \log \eps^2  +   2 \, \tr    Y_{21}^\dag Y_{21}  \log \eps^2  -   2 \, \tr Y_{21}^\dag Y_{21} \log X_\eps^2 
   ~  \longrightarrow  ~   -   2 \, \tr Y_{21}^\dag Y_{21} \log X^2 \, ,
\ee
 which is exactly what we found by considering $S(X^\dag X)$.

 Since $F^\dag P_{B^\perp} F = F^\dag F$, the  term \eqref{singc}  
  can be written as
  \bee
     \tr  ( F Y_{22}^\dag  + Y_{22} F^\dag )^2 \int_0^\infty  \frac{\eps^2}{(\eps^2 + u )^2} du  =    \tr  ( F Y_{22}^\dag  + Y_{22} F^\dag )^2 \\
  \eee   
 where we used  
 $\int_0^\infty  \frac{\eps^2}{(\eps^2 + u )^2} du  = - \eps^2 \tfrac{1}{\eps^2 + u } \Big|_0^\infty  = 1 $.
This  term does   not $\raw 0 $ as $\eps \raw 0 $.    However, since the term  \eqref{Reps} 
contains $- 2 \tr Y_{22} Y_{22}^\dag \log \eps^2 $ which diverges to $+ \infty$ we again  conclude that
 non-zero $Y_{22} $ must increase the second derivative (and, hence, the entropy).   
 This also proves Theorem~7 of \cite{GF} which says that $D_2[X,Y] $ diverges to $+ \infty $
 if and only if $Y_{22} $ is non-zero.
 
 \section{Extension to relative entropy} \label{sect:relent}

\subsection{Local additivity}

The relative entropy is defined (when $\ker \omega \subseteq \ker \rho $) as
\be
    H(\rho, \omega) =  \tr \rho (\log \rho - \log \omega)  = -S(\rho) - \tr \rho \log \omega
\ee
The arguments above are easily generalized to show that the maximization of relative entropy
with respect to a fixed reference state is locally additive because
        \be   \label{Hderiv1}
   \frac{d~}{dt} H \big( \rho(t), \omega  \big)     & = &  
             -  \frac{d~}{dt}  S[ \rho (t ) ] -  \tr   \rho^\prime(t) \log \omega
      \ee
         and
           \be   \label{Hderiv2}
             \frac{d^2~}{dt^2}  H\big(\rho (t), \omega\big)   & = & 
                    \frac{d^2~}{dt^2} S[ \rho (t ) ]  - \tr \rho^{\prime\prime} (t) \log \omega \,. 
       \ee                       
 Since the terms involving
$\log \omega$ are additive and the others come from differentiating $S(\rho)$, the following result  
is a straightforward corollary to our results for the minimal output entropy.

\begin{thm}  \label{thm:relent}
Let $\Phi_A :  M_{d_A} \mapsto  M_{d_B} $ and  $\Phi_C :  M_{d_C} \mapsto  M_{d_D} $  be quantum
channels, and let the states $\omega_A, \omega_C $ be fixed.
If   $  H\big[ \Phi_A(\rho_A) , \Phi_A(\omega_A) \big] $  and
 $   H\big[ \Phi_C(\rho_C) , \Phi_C(\omega_C) \big] $  have  non-degenerate  local maxima at
 $ \rho_A = \proj{\psi_A} $ and  $\rho_C  = \proj{\psi_C}  $, then
 $  H\big[( \Phi_A \ot \Phi_C)(\rho_{AC}) \, , \, ( \Phi_A \ot \Phi_C)(\omega_A \ot \omega_C) \big] $ 
 has a local maximum at $\rho_{AC} = \proj{ \psi_A \ot \psi_C} $.
\end{thm}

\subsection{Capacity of a quantum channel}

This result is of particular interest in studying the additivity of  Holevo capacity which describes the 
capacity of a quantum channel to transmit classical information using product inputs.  This is defined as
\be   \label{cholv}
    C_\hv(\Phi) = \sup_{ \pi_j, \rho_j } \Big(  S\big[ \Phi(\rho_\av)\big]  - \sum_j \pi_j S\big[ \Phi(\rho_j) \big] \Big)
            = \sup_{ \pi_j, \rho_j }  \chi(\{\pi_j, \Phi(\rho_j )\} ) 
\ee
where $\rho_\av = \sum_j \pi_j \rho_j $ and 
$  \chi(\{\pi_j, \rho_j \} ) =  S(\rho_\av) - \sum_j \pi_j S(\rho_j)$ and  $\rho_j = \proj{\psi_j} $ is a pure state. The strict concavity of $S(\rho)$ implies 
that the optimal output average $\Phi(\rho_\av$)  is unique.  

The optimization over ensembles can
be replaced by the following
 max-min expression  \cite{OPW,SW}   \be   \label{hvmaxmin}
    C_\hv(\Phi) & = &  \min_{\gamma } \max_{ \rho}  H  \big[  \Phi(  \rho) \, , \,   \Phi(\gamma)  \big]  \\
        & = & \max_{ \rho } H\big[   \Phi (\rho) \, , \,   \Phi(\rho_\av)  \big]   \label{mmav}
\ee
where $\rho, \gamma $ are density matrices. 
It follows from \eqref{mmav} that  the outputs $\Phi(\rho_j) $ in
any ensemble that optimizes \eqref{cholv} are ``equi-distant'' from the optimal average output $  \Phi(\rho_\av) $  since they satisfy
$H\big[   \Phi (\proj{\psi_j} ),  \Phi(\rho_\av)  \big]  =   C_\hv(\Phi)$.  It also follows from \eqref{mmav}  that
 $ H\big[  \Phi(\rho) \, , \,   \Phi(\rho_\av)  \big]  $ has a (possibly degenerate) local maximum at each  input  $\rho_j = \proj{\psi_j}$
 in an ensemble that maximizes \eqref{cholv}.
Moreover, this max-min expression gives an important criterion for superadditivity  which  we state only in the simplest case.

\begin{thm}   \label{thm:super}
A channel  $\Phi: M_d \mapsto M_d $ satisifes $C_\hv(\Phi \ot  \Phi) > 2 C_\hv(\Phi) $ if and only if 
there exists a  $| \Psi  \ket \in {\bf C}_d \ot {\bf C}_d$ such that 
\be     \label{supercond}
  H \big[  ( \Phi \ot \Phi) ( \proj{\Psi }) , (\Phi \ot \Phi)( \rho_\av \ot \rho_\av)  \big] > 2 \, C_\hv(\Phi).
\ee
\end{thm}
\noindent{\bf Proof:}  First, observe that superadditivity and \eqref{hvmaxmin} imply   
\bee   
       2 \, C_\hv(\Phi) <  C_\hv(\Phi \ot  \Phi) & = & 
         \max_{\Psi} H\big[ ( \Phi \ot \Phi) ( \proj{\Psi}), ( \Phi \ot \Phi) (\Gamma_\av ) \big] \nn \\
       &  \leq &    \max_{\Psi}  H\big[( \Phi \ot \Phi) (\proj{\Psi}) , (\Phi \ot \Phi)( \rho_\av \ot \rho_\av) \big]
\eee
where the max is taken over   $|\Psi \ket \in {\bf C}_d \ot {\bf C}_d$ and $\Gamma_\av$ 
is the true optimal average input for $\Phi \ot \Phi$.

To prove the converse recall that the optimal average output $(\Phi \ot \Phi) (\Gamma_\av) $ is unique and observe that one can 
always achieve   $ 2 \, C_\hv(\Phi) $ with a ``product ensemble'' 
 $  \chi\big( \{ \pi_j \pi_k,  (\Phi \ot \Phi)( \rho_j \ot \rho_k) \} \big) = 2 \, C_\hv(\Phi) $
 for which the optimal average input is  $  \rho_\av \ot \rho_\av$.  If this is not the true $\Gamma_\av$ 
then the supremum in \eqref{cholv} must be $ > 2 \, C_\hv(\Phi)$.  On the other hand if
$  \rho_\av \ot \rho_\av = \Gamma_\av $ then \eqref{mmav} and \eqref{supercond}  imply superadditivity.
\qed

\subsection{Implications for numerical work}

Theorem~\ref{thm:super}  implies that one can demonstrate
superadditivity  of the capacity without the need to find either the capacity $C_\hv(\Phi \ot \Phi)$ 
or the optimal average input $\Gamma_\av$.
It suffices to show that
  \eqref{supercond} holds.  If  $\rho_\av$ and $C_\hv(\Phi)$ for a single
use of the channel have been determined, the additional numerical effort is comparable to 
showing that  $S_{\min}(\Phi)$ is not additiive.

Before Hastings' work, Shor \cite{S} had shown the equivalence of several additivity conjectures
including those for both $C_\hv(\Phi) $ and $S_{\min}(\Phi) $.  Thus, Hastings' breakthrough
also implies superadditivity of $C_\hv(\Phi )$.  To find explicit examples of channels which violate
additivity, attention focussed on $S_{\min}(\Phi) $ as the simplest.  But subsequent work
suggests that violations sufficiently large to be observed requires very large dimensions. 
(See   \cite{CN} and endnotes to    \cite[Chapter~8]{AS}.)   It is, however,
plausible that examples for superadditivity of $C_\hv(\Phi) $ can be found in lower dimensions.  

If  $\proj{\psi_j} $ are optimal inputs for $H\Big[   \Phi (\rho) \, , \,   \Phi(\rho_\av)  \Big] $,
then  $ H\big[( \Phi \ot \Phi) ( \Gamma) , (\Phi \ot \Phi)( \rho_\av \ot \rho_\av)  \big] $ will always have local maxima at
$\Gamma = \proj{ \psi_j \ot \psi_k } $.  If  $C_\hv(\Phi )$ is superadditive, There will also be at least one 
entangled input $|  \Psi_{\max} \ket $ for which there is a local maximum satisfying \eqref{supercond}. 
Although finding this $|  \Psi_{\max} \ket $ might seem daunting in view of the many local maxima at
products  $|  \psi_j \ot \psi_k \ket $, that knowledge might also help to begin the search with highly 
entangled inputs designed to avoid those products.
 
 \bigskip

 \noindent  {\bf Acknowledgment:}  The work of JY  was supported in part by NSERC Discovery Grant RGPIN-2018-04742 as well as by Perimeter Institute for Theoretical Physics. Research at Perimeter Institute is supported by the Government of Canada through Innovation, Science and Economic Development Canada and by the Province of Ontario through the Ministry of Research, Innovation and Science.
The work of MBR was partially supported by NSF grant CCF 1018401 which was administered by Tufts University.
   Part of this work was done when MBR was visiting the Institute for Quantum Compuing in Waterloo, Canada and
   both authors were participating in  workshops at the Aspen Center for Theoretical Physics, Texas A\&M University and Guelph
   University.   The authors are grateful to an anonymous referee for a careful reading of an earlier version of the
   manuscript.

\appendix

     \section{Proof of key inequaity }  \label{app:phi}
     
       \numberwithin{equation}{section}
       
       \setcounter{equation}{0} 
     We present two proofs of the key inequality  \eqref{phikey2}, i.e., $ 2 \phi(ab) \leq \phi(a^2) + \phi(b^2) $
   with $\phi(a) = \frac{a+1}{a-1} \log a^2$ as in  \eqref{phidef}.  Before doing so, we make some observations useful in both arguments.
   
   a)    $a > 0$ implies $\phi(-a) \leq \phi(a) $ so that  it suffices to prove  \eqref{phikey2} for $a, b > 0 $.

   b)  It is straightforward to verify that $\phi(x)$  is continuous on    $(0,\infty)$ with a minimum at $x = 1$ satisfying  $\phi(1) = 4$. 

   c) $\phi(a) = \phi(a^{-1}) $  and equality holds in  \eqref{phikey2}, when $a = b$..  \medskip
 
  \noindent{\bf Elementary proof:}   We now assume $a, b > 0$ and observe that  \eqref{phikey2} is equivalent to
   \begin{align}
      \frac{ab + 1}{ab -1}\big[ \log a  + \log b  \big]   ~ &  \leq   ~\frac{a^2 + 1}{a^2 -1} \log a  + \frac{b^2 + 1}{b^2 -1} \log b   \nn  \\
      \intertext{which is equivalent to} 
  \frac{b-a}{ab-1}  \big[ \zeta(a)-  \zeta(b) \big] ~ &  \geq ~ 0    \qquad  \hbox{with}  \qquad  \zeta(x)  =  \frac{   x \log x}{ x^2 -1}   \, .   \label{zeta}
      \end{align}
Now observe that  $\zeta(x)  =  \zeta(x^{-1}) $ and
$   \zeta^\prime (x) = \frac{1}{x^2 -1 }  \Big(  1 - \frac{1}{4} \phi(x^2) \Big) $.  Thus, when $x > 1$, $\zeta^\prime(x) < 0 $ and
$\zeta$ is strictly decreasing for $x > 1$.  We can assume  $b > a$ so that it suffices to consider  three cases:

  $\bullet~$  When $ 1 < a < b $, this implies   $ \zeta(b)  <   \zeta(a)  $ so that \eqref{zeta} holds.
  
     $\bullet~$  When $a <  1 < b$ and $ab > 1$, then $ 1 < a^{-1} < b $ and $ \zeta(b) < \zeta( a^{-1}) = \zeta(a) $ so that \eqref{zeta} holds. 
      
      $\bullet~$   When $ a < 1 < b  $ and $ab < 1$,  then $1 <  b < a^{-1} $ so that  $\zeta(b) >  \zeta(a^{-1})  = \zeta(a) $ which 
      again implies \eqref{zeta}.

\noindent  Combining this with the continuity of $\phi(a) $ at $a =1 $ proves  \eqref{phikey2}. \qed  \medskip

\noindent{\bf Convexity Proof:}   
 Let $\chi(x) = \phi(e^x) = 2x \tfrac{e^x + 1}{e^x -1} $.   Then $\chi(\log a) = \phi(a)$ and  \eqref{phikey2} is equivalent to
\be  \label{chiconvex}
  \chi( \log ab) =    \chi \big( \half \log a^2 + \half \log b^2 ) \leq  \half  \chi( \log a^2  ) + \half \chi(\log b^2) 
     \ee
  which holds if $\chi(x)   $ is convex.  One can verify that
    \bee
     \chi^\prime(x) & = &  \frac{\sinh(x) - x}{ \sinh^2 (x/2) }  \\
 \chi^{\prime  \prime}(x)  & = &   \frac{x \coth(x/2) -2}{ \sinh^2 (x/2) } =  \frac{  \chi(x) - 4} {2 \sinh^2 (x/2) } \, .  \eee
Observation (b) above implies   $\chi(x) \geq \chi(0) = 4 $ on  $ {\bf R}$  which  implies
 that $\chi^{\prime  \prime}(x) \geq 0 $.    Thus $\chi(x) $ is convex and 
\eqref{chiconvex} holds.  \qed


\section{Derivative formulas involving $\log \rho(t)$}   \label{app:deriv}

\subsection{Derivative of  $\log \rho(t)$}  \label{app:D1log}

Let $\rho(t) $ 
be a one parameter family of density matrices  twice differentiable
in some neighborhood $\cN(0) $ of $ t = 0 $ so that for $t_1 \in \cN(0) $ 
there is a neighborhood $\cN(t_1)  \subset \cN(0)$ in which
\be
     \rho(t) = \rho(t_1) + (t - t_1) \rho^\prime(t_1) + O(t - t_1)^2
\ee
and $\rho(t) $ has full rank.
To find  the derivative of  $\log \rho(t) $ we begin with 
   the integral representation 
 \be  \label{intrep}
   \log \rho =   \lim_{M \raw \infty}   \int_0^M \Big(  \frac{1}{1 + u} -\frac{1}{ \rho + u  }  \Big) du 
   \ee
  which is valid when $\rho$ has full rank.
 Then 
    \be 
\lefteqn{    \log \rho(t_2) - \log  \rho(t_1) =   \int_0^\infty \Big( \frac{1}{\rho(t_1) + u I } -  \frac{1}{\rho(t_2) + u I } \Big)  du }  \qquad \qquad \nn \\
  & = &     \int_0^\infty  \frac{1}{\rho(t_1) + u I } \big[ \rho(t_2) - \rho(t_1) \big]  \frac{1}{\rho(t_2) + u I }  du  \nn \\
           & = &(t_2 - t_1)     \int_0^\infty  \frac{1}{\rho(t_1) + u I }   \rho^\prime(t_1)  \frac{1}{\rho(t_2) + u I }  du + O(t_2-t_1)^2  \nn \\
           & = &    (t_2 - t_1)     \int_0^\infty  \frac{1}{\rho(t_1) + u I }   \rho^\prime(t_1)  \frac{1}{\rho(t_1) + u I }  du  +  O(t_2-t_1)^2    
            \ee
     where we used $A^{-1} - B^{-1} =  A^{-1} (B-A) B^{-1} $.  Then
 \be  \label{logderiv}
     \frac{d~}{dt} \log \rho(t) \Big|_{t = t_1} & \equiv  &    \lim_{t_2 \raw t_1}   \frac{\log \rho(t_2) - \log  \rho(t_1)}{t_2 - t_1 } \nn \\
     & = &    
     \int_0^\infty    \frac{ 1}{\rho(t_1 ) + u I } \rho^\prime(t_1)  \frac{ 1}{\rho(t_1 ) + u I }  du 
     \ee

\subsection{ Identity  \eqref{BKM} for the modular operator  $\Delta_\rho =  L_\rho R_\rho^{-1}$}  \label{app:modop}
Next, observe that we can use \eqref{intrep} to show that 
\be  \label{preBKM}
   \big( L_{\log \rho} -   R_{\log \rho} \big) (\Gamma) & =  &  \int_0^\infty   \bigg[ \Gamma \frac{1}{\rho + uI} - \frac{1}{\rho + uI} \Gamma \bigg]  du   \nn \\
        & = &  \int_0^\infty  \frac{1}{\rho + uI}[ \rho  \Gamma  - \Gamma \rho ]\frac{1}{\rho + uI}      du  \nn \\
         & = & \big( L_\rho - R_\rho \big)  \int_0^\infty  \frac{1}{\rho + uI} \,  \Gamma\,  \frac{1}{\rho + uI}      du ~.
\ee
Then dividing by $ L_\rho - R_\rho$ and using $L_{\log \rho} = \log L_\rho$ gives \eqref{BKM} with $\Delta_\rho =  L_\rho R_\rho^{-1}$ .   

Since    $\frac{\log x}{x-1} $ can be defined by continuity at $x = 1$,
      $ \frac{ \log \Delta_\rho}{L_\rho - R_\rho } (W) = R_\rho^{-1}   \frac{ \log \Delta_\rho}{\Delta_\rho -  I}(W)   =  \rho^{-1}  W$ 
      when $L_\rho W  = R_\rho W$.

\subsection{ Bound on third derivative}   \label{app:D3bound}

Let $L(t) = \log \rho(t) $.  Then it follows immediately from  Eqs.  (5) and (8)  in Section 3.1   that 
\be
  -   \frac{d^3~}{dt^3} S[\rho(t)  ]  & = &   \tr \rho^{\prime \prime \prime} (t)  L(t)  + 2  \, \tr \rho^{\prime \prime }(t) L'(t)  
        + \tr \rho^\prime(t)  L^{\prime \prime } (t)   \\       
        & = &   \tr \rho^{\prime \prime \prime} (t) \log \rho(t) + 
         2 \tr \rho^{\prime \prime }(t)  \int_0^\infty    \frac{ 1}{\rho(t ) + u I } \rho^\prime(t)  \frac{ 1}{\rho(t ) + u I }   \, du \nn \\
     & ~ &     +   \tr \rho^{\prime}(t)   \frac{d~}{dt}  \int_0^\infty    \frac{ 1}{\rho(t ) + u I } \rho^\prime(t)  \frac{ 1}{\rho(t ) + u I }  \, du
    \label{deriv3}    \, .   \ee
When  $\rho(t) $ is given  by (9), $\rho(0) = X^2 $ is strictly positive definite.  Therefore, one can find an $\alpha > 0 $
and a $\tau \in (0,1) $ such that  $t \in (-\tau,\tau) $ implies that  $I_d > \rho(t) > \alpha_0 $, which implies that one 
can find $\{ A_k: k = 0, 1, 2\}$ such that
\be
    |  \tr   \log \rho(t)| < A_0   \qquad  \quad  \tr \rho^{-1} (t)  < A_1  \qquad  \quad 
    \tr \rho^{-2} (t) < A_2  
\ee
for all $t \in (- \tau,\tau) $.

Next, observe that  $  \tr X^2 = \tr Y Y^\dag = 1$,  implies $\| X \| = \tr |X| \leq 1 $ and   $\| Y \| = \tr |Y| \leq 1 $.  
Since the derivatives are given by 
\bsq \bee
     \rho^{(1)}(t)   ~ = ~   \rho^\prime(t) & = &  2 t (YY^\dag -  X^2) + \tfrac{1 - 2 t^2}{\sqrt{1 - t^2 } }(X Y^\dag + Y X )  \\
    \rho^{(2)}(t)   ~ = ~     \rho^{\prime \prime}(t)   & = &  2 (YY^\dag -  X^2)  +    \tfrac{    t(2t^3 -3) }{(1 - t^2)^{3/2}} (X Y^\dag + Y X ) \\
      \rho^{(3)}(t)   ~ = ~     \rho^{\prime  \prime  \prime}(t)   & = &      \tfrac{ -3   }{(1 - t^2)^{5/2}} (X Y^\dag + Y X ) 
\eee \esq
and  $0 < \tau < 1 $,   
one can find   $\{ B_k > 0 : k =   1, 2,3 \}$   such that   $|\rho^{(k)}(t)  |< B_k $ on $(-\tau  ,\tau ) $.  

 Then it is straightforward to
show that  there is an $R > 0 $ such that
\be
  \Big|   \frac{d^3~}{dt^3} S[\rho(t)  ]  \Big|  < R     \quad \forall ~ t \in (-\tau,\tau)  \, .
\ee
In particular, the  magnitudes of  the first and second terms in the third derivative are bounded by $B_3 A_0 $ and    $ 2 B_1 B_2  A_1 $ respectively,
while that for the third term is bounded by   $B_1 B_2 A_1 + 2 B_1^3 A_2  $.

  \pagebreak

\end{document}